\documentclass[10pt,conference,letterpaper]{IEEEtran}
\usepackage{graphicx}
\usepackage{epstopdf}
\usepackage[cmex10]{amsmath}
\usepackage{cases,enumerate}
\usepackage{subeqnarray,subfigure}
\usepackage{cite,setspace}
\usepackage{amsmath,bm,amssymb,amsthm}
\usepackage{amsfonts}
\usepackage{epsfig}
\usepackage{multirow}
\usepackage{amsmath,color}
\usepackage{caption}

\usepackage{times}
\usepackage[T1]{fontenc}
\usepackage[utf8]{inputenc}

\usepackage[linesnumbered,ruled,vlined]{algorithm2e}

\newtheorem{theorem}{\textbf{Theorem}}

\begin{document}

	\title{Distributionally Robust Optimization for Digital Twin Service Provisioning over Edge Computing}
	
	\author{ 
		\small 
		Yuxiang Li, 
		Jiayuan Chen
		and Changyan Yi\\
		\IEEEauthorblockA{\small College of Computer Science and Technology, Nanjing University of Aeronautics and Astronautics, Nanjing, China \\
			\text{\small Email: \{yuxiangli, jiayuan.chen, changyan.yi\}@nuaa.edu.cn}\\
		}
	}

	\IEEEtitleabstractindextext{%
	\begin{abstract}
		Digital Twin (DT) is a transformative technology poised to revolutionize a wide range of applications. This advancement has led to the emergence of digital twin as a service (DTaaS), enabling users to interact with DT models that accurately reflect the real-time status of their physical counterparts. Quality of DTaaS primarily depends on the freshness of DT data, which can be quantified by the age of information (AoI). The reliance on remote cloud servers solely for DTaaS provisioning presents significant challenges for latency-sensitive applications with strict AoI demands. Edge computing, as a promising paradigm, is expected to enable the AoI-aware provision of real-time DTaaS for users. 
		In this paper, we study the joint optimization of DT model deployment and DT model selection for DTaaS provisioning over edge computing, with the objective of maximizing the quality of DTaaS.
		To address the uncertainties of DT interactions imposed on DTaaS provisioning, we propose a novel distributionally robust optimization (DRO)-based approach, called Wasserstein DRO (WDRO), where we first reformulate the original problem to a robust optimization problem, with the objective of  maximizing the quality of DTaaS under the unforeseen extreme request conditions. Then, we leverage multi-level dual transformations based on Wasserstein distance to derive a robust solution. Simulations are conducted to evaluate the performance of the proposed WDRO, and the results demonstrate its superiority over counterparts.
	\end{abstract}
	}
	
	\maketitle
	\IEEEdisplaynontitleabstractindextext
	\IEEEpeerreviewmaketitle

	\section{Introduction}
	\IEEEPARstart{D}{igital} Twin (DT) is an advanced technology that aims to create highly accurate, dynamic DT models of physical entities (PT), serving as precise digital replicas that mirror the real-time status of their counterparts. 
	It is widely applied across fields such as intelligent manufacturing, autonomous driving, and personalized healthcare\cite{chen2024gene}.
	This advancement has led to the emergence of digital twin as a service (DTaaS), enabling users to interact with DT models.
	High-quality of DTaaS depends on the freshness of DT service response data, ensuring users to receive accurate and timely interaction experiences. The freshness of DT data can be quantified by the age of information (AoI).
	However, traditional remote cloud-based DTaaS provisioning, with DT models deployed on solely cloud servers, often fails to meet the demands of latency-sensitive applications, resulting in high AoI and a degradation in the quality of DTaaS\cite{fan2021digital}.
	Edge computing, as a promising paradigm, which deploys DT models on edge servers (ESs), may be an alternative to provide AoI-aware real-time DTaaS for users.
	
	Although some studies have focused on optimizing DT services over edge networks \cite{zhou2023hierar}\cite{shi2024service}, there are some critical issues which have not yet been well investigated. On one hand, the issue of updating DT models to preserve their accuracy has been overlooked, which leads to outdated data of service responses, and severely compromising quality of DTaaS. On the other hand, all of these studies often rely on the assumption that all DT interaction information is fully known in advance, which is impractical for DT with dynamic evolutions and uncertain access requests. 
	However, addressing these issues is very challenging because of the following reasons. i) The quality of DTaaS is influenced by both the fidelity of DT models and service latencies, but designing a metric that integrates these two elements is not straightforward due to their significantly heterogeneity.
	ii) Moreover, the uncertainty of the initiation of future DT interaction requests exacerbates the complexity of the seamlessly provisioning of DT services in the system optimization. 
	
	In this paper, we study a joint optimization of DT deployment and DT model selection for DTaaS provisioning over edge computing. Considering the enhancement of DTaaS provisioning on edge servers rather than remote cloud servers, we design a utility gain based on the difference of AoI as the metric to evaluate quality of DTaaS.
	We aim to maximize the total utility gain of all DT interaction requests by jointly determining: i) which DT models should be deployed on which ESs and ii) which DT models should be selected to serve which DT interaction requests, subjected to the storage resource constraints of ESs. Solving this problem is very challenging mainly due to the uncertainties of DT interaction requests and the resulted impacts to the DTaaS provisioning. 
	To address this, we propose a novel distributionally robust optimization (DRO)-based approach, called Wasserstein DRO (WDRO), where we employ the DRO method to reformulate the original problem, and then apply multi-level dual transformations based on Wasserstein distance to obtain a robust optimization solution.
	
	The main contributions of this paper are summarized in the following. 
	\begin{itemize}
		\item A joint optimization of DT model deployment and DT model selection for DTaaS provisioning over edge computing is first formulated, with the objective of maximizing the quality of DTaaS, measured by the total utility gain of all DT interaction requests.
		\item A novel DRO-based approach, namely WDRO, is proposed to obtain a robust solution. In WDRO, the original optimization problem is first reformulated employing the DRO method to account for uncertainty. Then, based on Wasserstein distance, multi-level dual transformations are applied to convert the problem into a form solvable by the Gurobi optimizer, resulting in a robust optimization solution that provides high quality of DTaaS under unforeseen extreme request conditions.
		\item Simulations are conducted to show the superiority of the proposed WDRO over counterparts.
	\end{itemize}
%

	\section{System Model and Problem Formulation}\label{SSMPF}
	\begin{figure}[!t]
		\centering
		\includegraphics[width=3.3in]{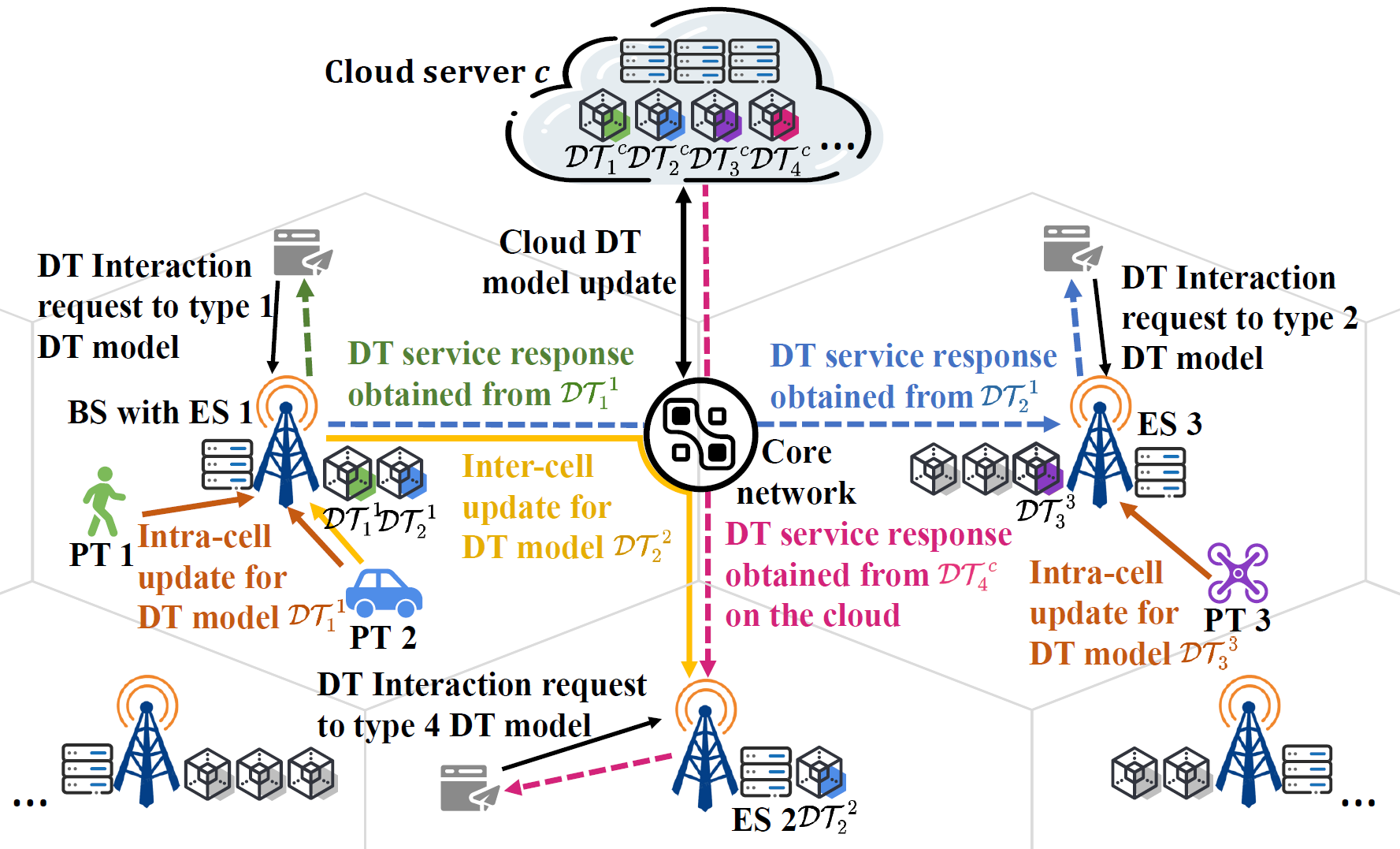}\\
		\caption{An illustration of the considered DTaaS provisioning over edge computing.}\label{model}
		\vspace{-0.7em}
	\end{figure}

	\subsection{System model} \label{SNM}
	
	Consider DTaaS provisioning over edge computing, as illustrated in Fig. \ref{model}, consisting of a set of geographically distributed edge servers (ESs) $\mathcal{V}$ with $|\mathcal{V}| = V$, each $v \in \mathcal{V}$ being associated with a base station, and a cloud server $c$. Each of them maintains interactive DT models for providing DT services for users. The system operates over continuous time, and we denote any time instant by $t$.
	There is a set of PTs $\mathcal{M}$ with $|\mathcal{M}| = M$, where each PT $m \in \mathcal{M}$ has a set of corresponding DT models, denoted as $\mathcal{H}_m = \left\{\mathcal{DT}^{g}_{m} \mid g \in \mathcal{V}_m \right\} \cup \left\{\mathcal{DT}^{c}_{m}\right\}$, where $\mathcal{V}_m$ is the subset of ES set $\mathcal{V}$ that maintain the DT models of PT $m \in \mathcal{M}$, $\mathcal{DT}^{g}_{m}$ represents the DT model of PT $m \in \mathcal{M}$ deployed on ES $g \in \mathcal{V}_m$, and $\mathcal{DT}^{c}_{m}$ represents the DT model of PT $m \in \mathcal{M}$ deployed on the cloud server $c$.
	For convenience, let $x_{m,v} = 1$ indicate that DT model of PT $m \in \mathcal{M}$ is deployed on ES $v \in \mathcal{V}$, $x_{m,v} = 0$ otherwise, and thereby subset $\mathcal{V}_m, m \in \mathcal{M}$ can be further expressed as $\mathcal{V}_m=\left\{v  \mid x_{m,v}=1, v \in \mathcal{V} \right\}$.
	
	To provide a high quality of DT services, the DT models within $\mathcal{H}_m$, $m \in \mathcal{M}$ 
	need to be periodically updated for	 keep synchronized with the corresponding PT $m \in \mathcal{M}$ by using data from PT $m \in \mathcal{M}$. 
	Similar to \cite{li2024AoI}, we adopt AoI to measure the performance of each DT model within $\mathcal{H}_m, m \in \mathcal{M}$ in synchronization with the corresponding PT $m \in \mathcal{M}$. We use $\mathcal{A}_{\mathcal{DT}_m^g}(t)$ and $\mathcal{A}_{\mathcal{DT}_m^c}(t), m \in \mathcal{M}, g \in \mathcal{V}_m$ to respectively represent the AoI of DT models maintained on ESs and the cloud server at time $t$.
	AoI is defined as the time elapsed since the most recent data for updating the DT model was transmitted from the PT. 
	Following this, at time $t$, the AoI of DT model $\mathcal{DT}^{g}_{m}(t), m \in \mathcal{M}$ maintained on ES $g \in \mathcal{V}_m$ can be expressed as $\mathcal{A}_{\mathcal{DT}^{g}_{m}}(t)=\mathcal{L}_{m,g}^{update}+(t-n \cdot \tau_m)$, where $n \in \left\{0,1,...,N\right\}$ represents the index of update period, determined as $n=\left\lfloor \frac{t}{\tau_m} \right\rfloor$, and $\tau_m$ is the time interval between two consecutive updates.
	$\mathcal{L}^{update}_{m, g}$ denotes the data transmission latency from PT $m \in \mathcal{M}$ to DT model $\mathcal{DT}_m^g$. 
	
\begin{figure}[!t]
	\centering
	\includegraphics[width=3.3in]{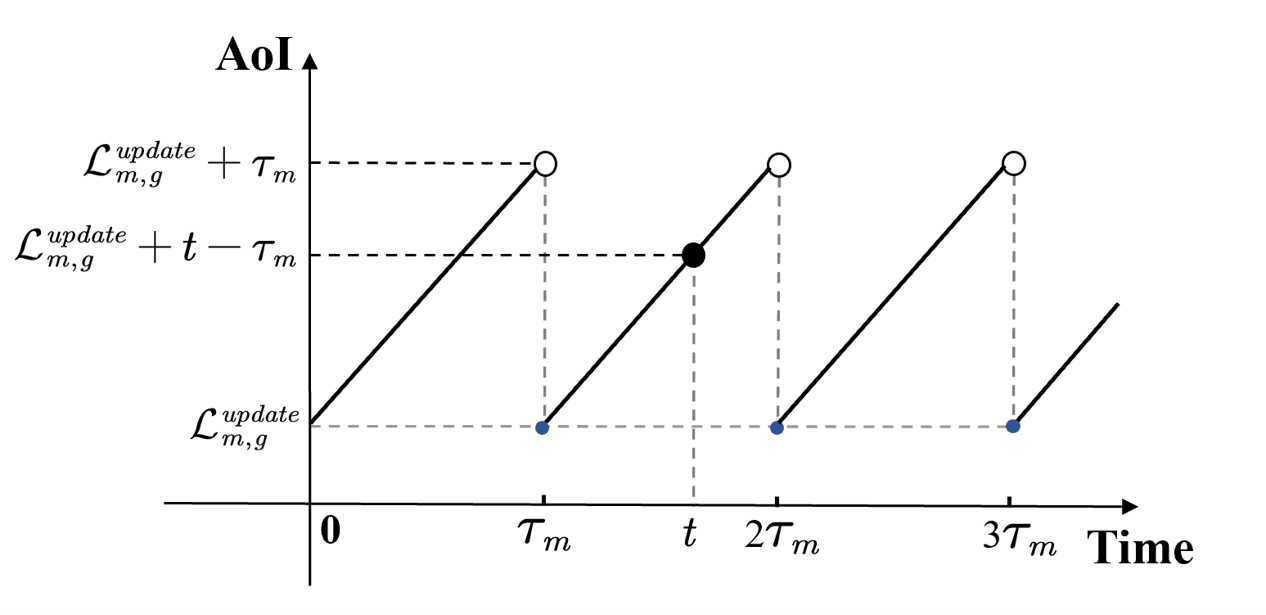}\\
	\caption{An illustration of the AoI evolution of DT model $\mathcal{DT}^{g}_{m}$.}\label{AoI}
	\vspace{-0.7em}
	\end{figure}

	Specifically, on the one hand, when $\mathcal{DT}^{g}_{m}, m \in \mathcal{M}, g \in \mathcal{V}_m$ and its corresponding PT $m \in \mathcal{M}$ are in the same cell, the data for updating DT model $\mathcal{DT}_m^g$ is transmitted directly from PT $m \in \mathcal{M}$ to ES $g \in \mathcal{V}_m$ (i.e., intra-cell update)\cite{chen2024three}. For intra-cell update, the transmission latency is expressed as $\mathcal{L}_{m,g}^{update}=a_{m} \cdot I^{upload}$, where $a_{m}$ denotes the size of data for updating from PT $m \in \mathcal{M}$, and $I^{upload}$ represents the latency of transmitting unit data to the local ES of PT $m \in \mathcal{M}$.
	On the other hand, when they are not in the same cell, the data for updating DT model $\mathcal{DT}_m^g$ is transmitted from PT $m \in \mathcal{M}$ to DT model $\mathcal{DT}_m^g, m \in \mathcal{M}, g \in \mathcal{V}_m$ through its local ES and the core network (i.e., inter-cell update). For inter-cell update, the transmission latency is given by $\mathcal{L}_{m,g}^{update}=a_{m} \cdot (I^{upload}+I_{loc_m,g}^{trans})$, where $loc_m \in \mathcal{V}$ indicates the coverage area of the ES in which PT $m \in \mathcal{M}$ is located, and $I_{loc_m,g}^{trans}$ represents the transmission latency of unit data between ES $loc_m \in \mathcal{V}$ and ES $g \in \mathcal{V}_m $ through the core network\cite{yi2023workload}. 
	We illustrate the AoI evolution of $\mathcal{DT}_m^g, m \in \mathcal{M}, g \in \mathcal{V}_m$, in Fig. \ref{AoI}.
	Similarly, at time $t$, the AoI of DT model $\mathcal{DT}_m^c, m \in \mathcal{M}$ maintained on the cloud server is calculated as $\mathcal{A}_{\mathcal{DT}_m^c }(t)=\mathcal{L}_{m,c}^{update}+(t-n \cdot \tau_m),  n \in \{0,1,...,N\}$, and $\mathcal{L}_{m,c}^{update}=a_{m} \cdot I_{loc_m,c}^{trans}$, where $I_{loc_{m},c}^{trans}$ denotes the latency of transmitting unit data from ES $loc_{m} \in \mathcal{V}$ to cloud server $c$.
	
	Users can initiate interaction requests to DT models at any time through their local ESs, and then receive the DT service responses from intra-cell, inter-cell ESs, or the cloud server\cite{yang2024dynamic}.
	We denote the set $\mathcal{R}$ as all DT interaction requests initiated by all users, and we represent each DT interaction request $r\in \mathcal R$ as a tuple $r=<t_r,loc_r,m_r,s_r^m>$, where $t_r$ is initiation time, and $loc_r \in \mathcal{V}$ is initiation location. Requests initiated from a specific location may exhibit preferences for certain types $m_r \in \mathcal{M}$. Particularly, $m_r \in \mathcal{M}$ denotes the type of DT model that the DT interaction request $r \in \mathcal{R}$ intends to interact with, where the DT service is provided by one of the DT models within $\mathcal{H}_{m}$, and $s_r^m$ is the data size of DT service response obtained from the DT model within $\mathcal{H}_m$. We denote the latency of DT service response for DT interaction request $r \in \mathcal R$ served by DT model $\mathcal{DT}^{g}_{m_r}, g \in \mathcal{V}_{m_r}$ as $\mathcal{L}^{response}_{g, r}=s_{r}\cdot I_{g,loc_{r}}$, where $I_{g,loc_{r}}$ represents the transmission latency of unit data between ES $g \in \mathcal{V}_{m_r}$ and $loc_r \in \mathcal{V}$.
	We use $y_{r,g}=1$ to signify that the DT model $\mathcal{DT}_{m_r}^g, g \in \mathcal{V}_{m_r}$ is selected to serve the DT interaction request $r\in \mathcal R$, and $y_{r,g}=0$, otherwise.
	Based on these, we define the AoI of DT service response to DT interaction request $r \in \mathcal{R}$ served by $\mathcal{DT}_{m_r}^g, g \in \mathcal{V}_{m_r}$ as $\mathcal{A}_r^g=\sum_{g\in\mathcal{V}_{m_r}}y_{r,g} \cdot (\mathcal{A}_{\mathcal{DT}^{g}_{m_r}}(t_r)+\mathcal{L}_{g,r}^{response})$.
	Particularly, if there is no DT model on ESs can provide the DT service for DT interaction request $r \in \mathcal{R}$, it can only obtain the DT service response from the DT model deployed on the cloud server $c$ with AoI $\mathcal{A}_r^c=\mathcal{A}_{\mathcal{DT}^{c}_{m_r}}(t_r)+\mathcal{L}_{c,r}^{response}$.
	It should be noted that, DT interaction request $r \in \mathcal{R}$ is uncertain, including its initiation time $t_r$, location $loc_r$, target type DT model $m_r$ and the size of DT service response $s_r^m$.

	Due to the long transmission latency between users and the cloud server, DT interaction requests served by DT models maintained on the cloud server results in high AoIs for the DT interaction requests\cite{li2024AoI}. In contrast, DT models maintained on ESs can provide higher-quality DT services with lower AoIs. We use the utility gain as a metric to evaluate the quality of DT service achieved through edge computing.
	Following this, the utility gain of DT interaction request $r \in \mathcal{R}$ served by $\mathcal{DT}_{m_r}^g, g \in \mathcal{V}_{m_r}$ is $u_{r}=\sum_{g\in\mathcal{V}_{m_r}}y_{r,g} \cdot (\mathcal{A}_r^c-\mathcal{A}_r^g)$.
	
	\subsection{Problem Formulation} \label{PF}
	With the objective of maximizing the total utility gain of all DT interaction requests within $\mathcal{R}$ while satisfying the storage resource constraints of each ES $v \in \mathcal{V}$, the joint optimization of DT model deployment $ \mathbf{X}=\{x_{m,v}\}_{\forall m \in \mathcal{M}, v \in \mathcal{V}}$ and DT model selection $\mathbf{Y} = \{y_{r,g}\}_{\forall r \in \mathcal{R}, g \in \mathcal{V}_{m_r}}$ can be formulated as
	\begin{subequations}
		\begin{align}
			& \mathcal{P}_1:\max_{\mathbf{X,Y}}\sum\nolimits_{r\in\mathcal{R}}\sum\nolimits_{g\in\mathcal{V}_{m_r}}y_{r,g} \cdot (\mathcal{A}_r^c-\mathcal{A}_r^g)\tag{\theequation}\\
			s.t. ~ &\sum\nolimits_{m\in \mathcal{M}}c_m\cdot x_{m,v}\leq \Phi_v, \: \forall v\in \mathcal{V}, \label{1}\\
			&\sum\nolimits_{g\in \mathcal{V}_{m_r}}y_{r,g}=1,\: \forall r\in \mathcal{R}, \label{5}\\
			&\mathcal{V}_m=\left\{v  \mid x_{m,v}=1, v \in \mathcal{V} \right\},\: \forall m \in \mathcal{M}, \label{v_m}\\
			&y_{r,g}\in\{0,1\},\: \forall r\in \mathcal{R}, g\in \mathcal{V}_{m_r}, \label{cons6}\\
			&x_{m,v}\in\{0,1\},\: \forall m\in \mathcal{M}, v\in \mathcal{V}, \label{cons7}
		\end{align}
	\end{subequations}
	where constraint (\ref{1}) is the storage constraint, and $c_m$ denotes the storage resource required to maintain a DT model of PT $m \in \mathcal{M}$, and $\Phi_v$ denotes the maximum storage capacity of the ES $v \in \mathcal{V}$; 
	constraint (\ref{5}) indicates that each DT interaction request $r\in \mathcal{R}$ can only be served by a single DT model $\mathcal{DT}_{m_r}^g, g\in \mathcal{V}_{m_r}$. 
	It is obvious that solving $\mathcal{P}_1$ requires prior knowledge of DT interaction requests, which is unavailable due to the uncertainty of all future DT interaction requests within $\mathcal{R}$. 
	Intuitively, historical interaction information can provide valuable insights into future uncertain interactions to some extent. However, most existing approaches heavily rely on these historical information to guide solution generation, making them susceptible to overfitting and unable to effectively handle unforeseen extreme request conditions, resulting in the degradation of total utility gain of all DT interaction requests. To address this, we propose a DRO-based approach that aims to derive a robust solution to maintain a high total utility gain of all DT interaction requests, even under the unforeseen extreme request conditions, namely the distribution of future uncertain DT interaction requests significantly deviates from historical ones. 
	
	\section{DRO-based optimization algorithm}\label{SOLUTION}
	In this section, we proposed a DRO-based approach, called Wasserstein DRO (WDRO), to obtain a robust solution. Specifically, we first reformulate $\mathcal{P}_1$ employing the DRO method with a confidence set, and then apply multi-level dual transformations based on Wasserstein distance to make it solvable by the Gurobi optimizer.
	\subsection{DRO-based Problem Reformulation} \label{PRF}
	According to DRO\cite{erick2010Distributionally}, we first design a sample space based on all possible DT interaction requests. Based on tolerance values and the reference distribution derived from historical DT interaction requests, we construct a confidence set containing the true distribution with a certain confidence level. Based on these, we reformulate $\mathcal{P}_1$ into a min-max problem, aiming to optimize the utility gain under the unforeseen extreme request conditions that significantly deviates from historical ones.
	
	\textit{\textbf{Design the sample space $\Omega$}}: For each uncertain DT interaction request $r=<t_r,loc_r,m_r,s_r^m>, r \in \mathcal{R}$, the uncertain initiation time $t_r$ is cancelled within the calculation of utility gain, i.e., $\mathcal{A}_r^c-\mathcal{A}_r^g=\mathcal{A}_{\mathcal{DT}^{c}_{m_r}}(t_r)+\mathcal{L}_{c,r}^{response}-(\mathcal{A}_{\mathcal{DT}^{g}_{m_r}}(t_r)+\mathcal{L}_{g,r}^{response})=\mathcal{L}_{{m_r},c}^{update}-\mathcal{L}_{{m_r},g}^{update}+\mathcal{L}_{c,r}^{response}-\mathcal{L}_{g,r}^{response}$, and the uncertain data size of DT service response $s_r^m$ is predetermined by target type $m_r$. Therefore, the uncertain elements of each DT interaction request $r \in \mathcal{R}$ are actually the request initiation location $loc_r \in \mathcal{V}$ and the target type of DT model $m_r \in \mathcal{M}$. The sample space $\Omega$, denoted as $\Omega=\{e_1,e_2,...,e_k,...,e_K\}$ with $K=|\mathcal{V}|*|\mathcal{M}|$ sample points, contains all possible combinations of initiation locations and target DT model types. Each sample point $e_k \in \Omega$ represents a basic event, namely an DT interaction request that is initiated at a certain location and targets a specific DT model type. 
	
	\textit{\textbf{Construct the confidence set $\mathcal{D}$}}:
	We denote the set of historical DT interaction request as $\mathcal{R}^{\prime}$. For the sample space $\Omega$ with $K$ sample point, we design reference distribution $\mathbb{P}_0=\{p_1^0,p_2^0,...,p_K^0\}$ based on $\mathcal{R}^{\prime}$. 
	Additionally, we set the form of ambiguity distribution of future uncertain DT interaction requests as $\mathbb{P}=\{p_1,p_2,...,p_K\}$, which represents the unknown true distribution of future uncertain DT interaction requests.
	We use distance metric $d(\mathbb{P},\mathbb{P}_0)$ to represent the distance between $\mathbb{P}_0$ and the ambiguity distribution $\mathbb{P}$.
	Then, we construct a confidence set $\mathcal{D}$ for the ambiguity distribution $\mathbb{P}$, which can be expressed as
	\begin{equation}
		\mathcal{D}=\{\mathbb{P}:d(\mathbb{P},\mathbb{P}_0)\leq\theta\},
	\end{equation}
	where $\theta$ is the tolerance parameter indicating the maximum distance between the ambiguity distribution $\mathbb{P}$ and the reference distribution $\mathbb{P}_0$, and determines the size of the confidence set $\mathcal{D}$.
	Affected by the volume of historical DT interaction request $|\mathcal{R}^{\prime}|$ and confidence level $\beta \in [0,1]$, the tolerance parameter $\theta$ is calculated as $\theta=K\sqrt{\frac{2}{|\mathcal{R}^{\prime}|}\ln\frac{1} {1-\beta}}$\cite{30}.
	

	Additionally, we construct the reference distribution $\mathbb{P}_0$ using a widely recognized step function, which increases by $1/|\mathcal{R}^{\prime}|$ at each historical DT interaction request. Specifically, we define $p_k^0=\frac1{|\mathcal{R}^{\prime}|}\sum_{f=1}^{|\mathcal{R}^{\prime}|}\delta_{j}(k), k \in \{1,...,K\}$, where $\delta_{f}(k)=1$ if the $f$ data sample matches the basic event $e_k \in \Omega$, and $\delta_{f}(k)=0$ otherwise.

	\textit{\textbf{Reformulate the problem}}:
	According to DRO, instead of focusing on the unique true distribution, we build the confidence set $\mathbb{P}$ of the ambiguous distribution such that the true distribution exists with a certain confidence level in this set. 
	Considering risk aversion, we focus on the worst case distribution within the confidence set $\mathcal{D}$, ensuring robustness of DTaaS. To facilitate solution derivation, we convert the maximization objective into the minimization of its negative. Then we minimize the negative of the objective under the worst case in $\mathbb{P}$ under storage resource constraints. Following this, we reformulate $\mathcal{P}_1$ as a DRO problem
	\begin{subequations}
	\begin{align}
		& \mathcal{P}_2:\min_{\mathbf{X,Y}}\max_{\mathbb{P}\in \mathcal{D}}\sum\nolimits_{r\in\mathcal{R}}\sum\nolimits_{g\in\mathcal{V}_{m_r}}y_{r,g} \cdot (\mathcal{A}_r^g-\mathcal{A}_r^c)\tag{\theequation}\\
		s.t. ~ &(\ref{1})- (\ref{cons7}),\\
		&\mathcal{D}=\{\mathbb{P}:d(\mathbb{P},\mathbb{P}_0)\leq\theta\}.
	\end{align}
	\end{subequations}
	
	\subsection{Multi-Level Dual Transformation based on Wasserstein Distance} \label{MTM_WD}
	The min-max format of $\mathcal{P}_2$ motivates us to transform the internal maximization problem into a convex problem, so that it can be combined with the external minimization problem through dual transformations. 
	The process involves aggregating DT interaction requests and multi-level dual transformations based on Wasserstein distance\cite{Gao2016Distributionally}, which facilitates the derivation of a robust solution.
%
	

	\textit{\textbf{DT Interaction Request Aggregation:}}
	We divide the DT interaction requests $r \in \mathcal{R}$ into $K$ subsets $\mathcal {R}_1, \mathcal {R}_2,..., \mathcal {R}_K$ according to initiation location $loc_r$ and type of target DT model $m_r$, assuming that DT interaction requests within the same subset are served by the same DT model. Based on the definition of $u_r$, DT interaction request $r \in \mathcal{R}$ in the subset $R_k$ have the same $u_r$. Then, we can get
	\begin{equation}
		\begin{aligned}
			&\sum\nolimits_{r\in\mathcal{R}}\sum\nolimits\nolimits_{g\in\mathcal{V}_{m_r}}y_{r,g} \cdot (\mathcal{A}_r^g-\mathcal{A}_r^c) \\
			&=\sum\nolimits_{k=1}^{K}\sum\nolimits_{r\in \mathcal{R}_k}\sum\nolimits_{g\in \mathcal{V}_{m_r}} y_{r,g}\cdot (\mathcal{A}_r^g-\mathcal{A}_r^c)\\
			&=\sum\nolimits_{k=1}^{K}|\mathcal{R}_k|\cdot\sum\nolimits_{g\in \mathcal{V}_{m_r}} y_{r_k,g}\cdot (\mathcal{A}_{r_k}^g-\mathcal{A}_{r_k}^c)\\
			&=\sum\nolimits_{k=1}^{K}p_k\cdot\sum\nolimits_{g\in \mathcal{V}_{m_r}} |\mathcal{R}| \cdot y_{r_k,g}\cdot (\mathcal{A}_{r_k}^g-\mathcal{A}_{r_k}^c)\\
			&=\sum\nolimits_{k=1}^{K}p_k\cdot\Psi[X,Y,r_k],
		\end{aligned}
	\end{equation}
	where $r_k$ represents any DT interaction request within $\mathcal{R}_k, k \in \{1,...,K\}$, and $p_k$ is the ratio of the size of subset $R_k$ to the size of set $R$.

	\textit{\textbf{Problem-Level Dual Transformation Based on Wasserstein distance:}}
	We assume that $\mathbf{X}$ and $\mathbf{Y}$ are known in the external minimization problem, and focus on the internal maximization problem, which can be reformulated as
	\begin{subequations}
		\begin{align}
			&\max_{\mathbb{P}\in \mathcal{D}}\sum\nolimits\nolimits_{k=1}^{K}p_k\cdot \Psi[\mathbf{X},\mathbf{Y},r_k]\tag{\theequation}\\
			s.t. ~ &\sum\nolimits\nolimits_{k=1}^Kp_k=1,\\
			& p_k\geq0, \forall k \in \{1,...,K\}, \\	&\mathcal{D}=\{\mathbb{P}:d(\mathbb{P},\mathbb{P}_0)\leq\theta\}.\label{DistributeDistance}
		\end{align}
	\end{subequations}
	Then based on the definition of the Wasserstein distance, constraint (\ref{DistributeDistance}) can be transformed into
	\begin{subequations}
		\begin{align}
			 &\min_{\pi\geq0}\sum\nolimits\nolimits_{k,j=1}^{K}\|\xi_{k}-\xi_{j}\|\pi_{k,j}\leq\theta\tag{\theequation}\\& s.t.\sum\nolimits_{k=1}^{K}\pi_{k,j}=p_j^0,\forall j \in \{1,...,K\},\label{6b}\\
			&\qquad\sum\nolimits_{j=1}^{K}\pi_{k,j}=p_k,\forall k \in \{1,...,K\}, \label{Known}
		\end{align}
	\end{subequations}
	where $\xi_j=(\mathcal{T}_j, \mathcal{S}_j)^T$ denotes the average response time and data size of historical service response within subset $\mathcal{R}_j^{\prime}, j \in \{1,...,K\}$. In contrast, $\xi_k=(\mathcal{T}_k, \mathcal{S}_k)^T$ represents the corresponding metrics for real DT interaction requests within $\mathcal{R}_k, k \in \{1,...,K\}$. $||\xi_{k}-\xi_{j}||$ represents the distance between two sample points $e_{k}$ and $e_{j}$. $\pi_{k,j}$ represents the amount of mass transported from $\xi_{j}$ to $\xi_k$. Then we replace $p_k$ with equality constraints (\ref{Known}) to completely eliminate $p_k$. The internal maximization problem is further transformed into
	\begin{subequations}
		\begin{align}
			&\max_{\pi\geq0}\sum\nolimits_{k,j=1}^{K}\pi_{k,j}\cdot \Psi[\mathbf{X},\mathbf{Y},r_k]\tag{\theequation}\\
			s.t. ~ 	&\text{(\ref{6b})}, \notag\\
			&\sum\nolimits_{k,j=1}^{K}\|\xi_{k}-\xi_{j}\|\pi_{k,j}\leq\theta.
		\end{align}
	\end{subequations}
	Obviously, the objective is a linear combination in terms of $\pi$, and the constraints are linear inequalities or equations with respect to the variables. This is a convex linear programming problem, which can be dual transformed as
	\begin{subequations}
		\begin{align}
			&\min_{\lambda\geq0,h_{j}}\lambda\theta+\sum\nolimits_{j=1}^{K}p_j^0\cdot h_{j}\tag{\theequation}\\
			s.t. ~ & h_{j}+\lambda\|\xi_{k}-\xi_{j}\|\geq \Psi[\mathbf{X},\mathbf{Y},r_k], \forall j,k \in \{1,...,K\},  \label{p3}
		\end{align}
	\end{subequations}
	where $\lambda$ and $h_{j}$ is the dual variable introduced.

	\textit{\textbf{Norm-Level Dual Transformation for Constraint Optimization:}}	
	Considering the constraints (\ref{p3}) after dual transformation are typical discrete constraints. The complexity of the solution increases as the number of $j$ and $k$ combinations increases, which lead to computational difficulties. 
	To this end, we introduce the set $\Xi$ and the maximization operation, which allow us to partially de-discretize the problem and simplify the solution process.
	Specifically, $\xi_k$ is an unknown numerical characteristics vector of real DT interaction requests, and its value is continuous and uncertain. In order to describe this uncertainty, we refer to the continuous set $\Xi=\{\xi|c\cdot\xi \leq d\}$ to describe possible $\xi_k$, which makes the problem shift towards dealing with continuous uncertainty.
	We can transform the constraints (\ref{p3}) as
	\begin{equation}
		\begin{aligned}
			h_{j}\geq \max_{\xi\in \Xi}\Psi[\mathbf{X},\mathbf{Y},r_k]-\lambda\|\xi-\xi_{j}\|,\forall j,k \in \{1,...,K\}.
		\end{aligned}
	\end{equation}
	Then, express the norm in the formula by the dual of the dual norm, and we further transform the constraint(\ref{p3}) as
	\begin{equation}
		\begin{aligned}
			h_{j}\geq \max_{\xi\in \Xi}\Psi[\mathbf{X},\mathbf{Y},r_k]\hspace{-0.5mm}-\hspace{-2mm}\max_{\|n_{j}\|_*\leq\lambda} \hspace{-2mm} \|\xi-\xi_{j}\|^Tn_{j},\forall j,k \in \hspace{-1mm}\{1,...,K\} \label{maxmax}.
		\end{aligned}
	\end{equation}
	\begin{theorem}\label{dual-dual}
		norm $\|x\|$ is the dual of its dual norm.
	\end{theorem}
	\begin{IEEEproof}
		This proof is omitted due to the page limit.
	\end{IEEEproof}
	\begin{theorem}\label{dual}
		$\lambda\|\xi-\xi_{j}\|$ can be expressed as  $\max_{\|n_{j}\|_*\leq\lambda} \|\xi-\xi_{j}\|^Tn_{j}$.
	\end{theorem}
	\begin{IEEEproof}
		According to Theorem \ref{dual-dual}, for any vector \(x\), we have
		\[\|x\| = \max_{\|n\|_* \leq 1} |x|^T n.\]
		Therefore, for the vector \(\xi - \xi_{j}\), we have
		\[\|\xi - \xi_{j}\| = \max_{\|n_{j}\|_* \leq 1} \|\xi - \xi_{j}\|^T n_{j}.\]
		Such equivalence relation can be illustrated as
		\begin{align*}
			\max_{\|n_{j}\|_* \leq \lambda} \|\xi - \xi_{j}\|^T n_{j} & = \max_{\|n_{j}\|_* \leq 1} \|\xi - \xi_{j}\|^T (\lambda n_{j}) \\
			& = \lambda \max_{\|n_{j}\|_* \leq 1} \|\xi - \xi_{j}\|^T n_{j}.
		\end{align*}
		Combining the previous results, we have
		\[\max_{\|n_{j}\|_* \leq \lambda} \|\xi - \xi_{j}\|^T n_{j} = \lambda \|\xi - \xi_{j}\|.\]
		Therefore, we can get
		\[\lambda \|\xi - \xi_{j}\| = \max_{\|n_{j}\|_* \leq \lambda} \|\xi - \xi_{j}\|^T n_{j},\]
		which proves that \(\lambda \|\xi - \xi_{j}\|\) can be expressed as  \(\max_{\|n_{j}\|_* \leq \lambda} \|\xi - \xi_{j}\|^T n_{j}\).
	\end{IEEEproof}
	
	\textit{\textbf{Constraint-Level Dual Transformation and Order Exchange:}}	
	The right side of the equation (\ref{maxmax}) can be transformed based on the minimax principle, which leads to $\max_{\xi\in \Xi}\min_{\|n_{j}\|_*\leq\lambda} \Psi[\mathbf{X},\mathbf{Y},r_k]-\|\xi-\xi_{j}\|^Tn_{j}$. 

	We exchange the order of solving max-min problems, the inner layer maximization problem is transformed as
	\begin{subequations}
		\begin{align}
			&\min_{\lambda\geq0,h_{j},\|n_{j}\|_*\leq\lambda}\lambda\theta+\sum\nolimits_{j=1}^{K}p_j^0\cdot h_{j}\tag{\theequation}\\
			s.t. ~ &h_{j}\geq \max_{\xi\in \Xi}\Psi[\mathbf{X},\mathbf{Y},r_k]\hspace{-1mm}-\hspace{-1mm}\|\xi-\xi_{j}\|^Tn_{j},\forall j,k \in \{1,...,K\}, \label{d1}\\
			&c\cdot\xi \leq d.  \label{p4}
		\end{align}
	\end{subequations}
		\begin{theorem}\label{max-min}
		$\min_{\|n_{j}\|_*\leq\lambda} \max_{\xi\in \Xi}\Psi[\mathbf{X},\mathbf{Y},r_k]-\|\xi-\xi_{j}\|^Tn_{j}$ and $\max_{\xi\in \Xi}\min_{\|n_{j}\|_*\leq\lambda} \Psi[\mathbf{X},\mathbf{Y},r_k]-\|\xi-\xi_{j}\|^Tn_{j}$ are equivalent.
	\end{theorem}
	\begin{IEEEproof}
		This proof is omitted due to the page limit.
	\end{IEEEproof}
	We transform the constraints (\ref{d1}) and (\ref{p4}), obtaining a dual form as
	\begin{subequations}
		\begin{align}
			&n_{j}^{T}\xi_{j}+\Psi[\mathbf{X},\mathbf{Y},r_k]+z_{j}^{T}d \leq h_{j}, \forall j,k \in \{1,...,K\}, \label{a}\\
			&z_{j}^{T}d = n_{j},\forall j \in \{1,...,K\}, \\
			&\lambda,z_{j}\geq0,\forall j \in \{1,...,K\}, \\
			&\|n_{j}\|_*\leq\lambda,\forall j \in \{1,...,K\}.\label{b}
		\end{align}
	\end{subequations}
	After completing the dual transformation of the internal maximization problem, we combine it with the external minimization problem, and eventually reformulate $\mathcal{P}_2$ as
	\begin{subequations}
	\begin{align}
		&\mathcal{P}_3:\min_{\lambda,h_{j},n_{j} ,z_{j},\mathbf{X,Y}}\lambda\theta+\sum\nolimits_{j=1}^{K}p_j^0\cdot h_{j}\tag{\theequation}\\
		s.t. ~ &(\ref{1})- (\ref{cons7}), (\ref{a})-(\ref{b}).\label{que}
	\end{align}
	\end{subequations}
	$\mathcal{P}_3$ is a mixed-integer nonlinear programming problem characterized by the dual norm and the presence of both continuous and binary variables, making it NP-hard. 
	To effectively handle the mixed-integer programming and nonlinear components in the objective function and constraints, we employ the Gurobi optimization solver to solve the $\mathcal{P}_3$. 
%
%
	\vspace{-0.3em}
	\section{Simulation Results}\label{PE}
	\begin{figure*}[htbp]
		\begin{minipage}{0.30\textwidth}
			\centering
			\includegraphics[width=2.5in]{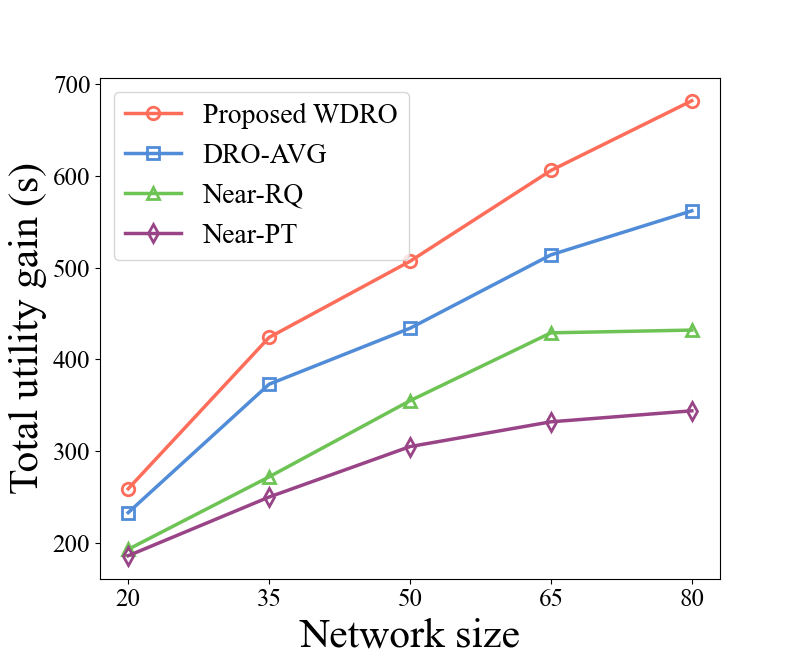}
			\captionof{figure}{Comparison of total utility gain w.r.t network size.}
			\vfill
			\label{networksize}
			\vspace{-0.3em}
		\end{minipage}\hfill
		\begin{minipage}{0.30\textwidth}
			\centering
			\includegraphics[width=2.5in]{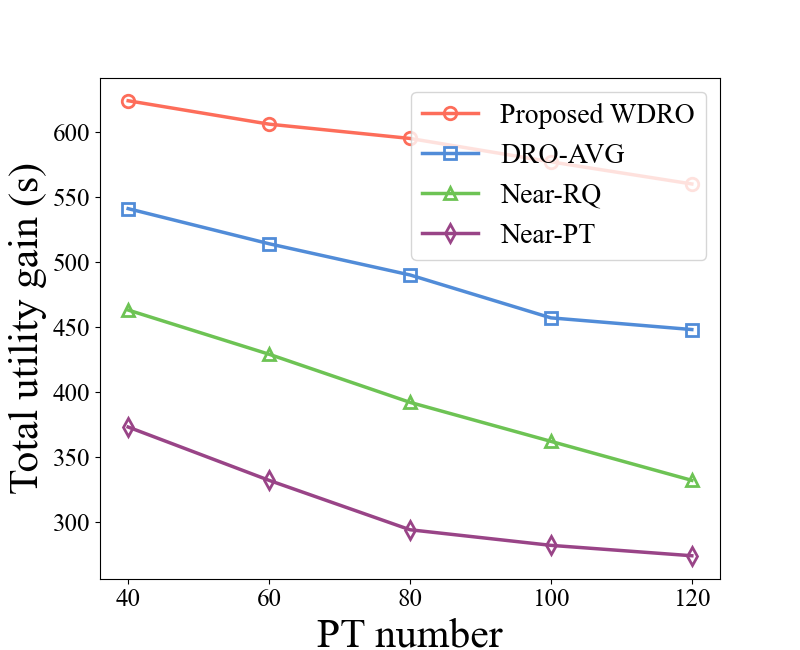}
			\captionof{figure}{Comparison of total utility gain w.r.t PT number.}
			\vfill
			\label{PT_number}   
			\vspace{-0.3em}
		\end{minipage}\hfill
		\begin{minipage}{0.339\textwidth}
			\centering
			\includegraphics[width=2.85in]{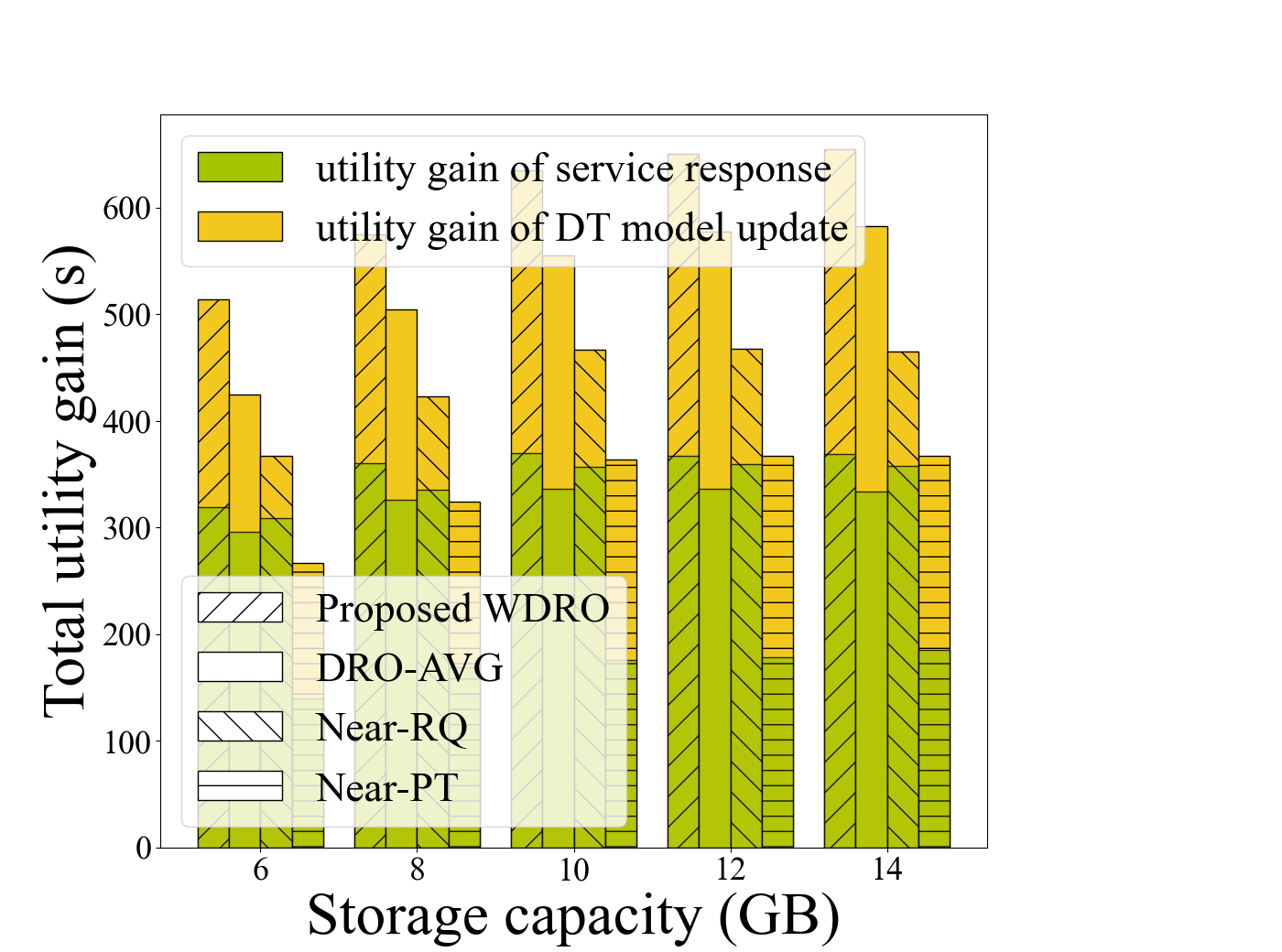}
			\captionof{figure}{Comparison of total utility gain w.r.t storage capacity.}
			\vfill
			\label{storage_capacity}
			\vspace{-0.3em}
		\end{minipage}
		\vspace{-1.5em}
	\end{figure*}
	Consider DTaaS provisioning over edge computing, the storage capacity of each ES is set within [8, 16] GB randomly, and the size of DT model $s_m, m \in \mathcal{M}$ is set within [0.5, 2] GB. The size of data for updating of PT $m \in \mathcal{M}$ is set within [2, 5] MB, while the size of the DT interaction request response data is set within [0.5, 2] MB. 
	Additionally, the transmission latency for sending a unit of data between two ESs is set within [0.2, 1] ms. The transmission latency for sending a unit data between the remote cloud and a ES through the core network is set within [2, 10] ms. We use the real-world request dataset from edge networks in \cite{dataset}, and divided them into historical DT interaction requests for constructing the reference distribution and future DT interaction requests.
	
	For the comparison purpose, the following schemes are simulated as benchmarks. 1) Near-PT: Prioritize the deployment of the DT model on the ES where the corresponding PT is located, and ignore the uncertainty of DT interaction request;
	2) Near-RQ: Deploy DT models based on historical DT interaction information, targeting the ES in the area with the highest visits, and ignore the update of DT models;
	3) DRO-AVG: Utilize a DRO-based approach to consider update of DT models and DT interaction requests, but use average distribution as the reference distribution.

	
	Fig. \ref{networksize} shows the total utility gain of different algorithms under different network sizes, thus demonstrating the applicability of the WDRO in large-scale networks. The results demonstrate that the proposed WDRO performs significantly better than the other three algorithms, and the total utility gain shows a clear upward trend as the network size increases. Compared with DRO-AVG, our algorithm exhibits a greater increase, which is due to the fact that our WDRO takes into account the similarity between the historical interaction information and the upcoming DT interaction requests. In contrast, Near-RQ and Near-PT exhibit slower increases, as they focus only on DT model updates or DT service response without optimizing overall DT service quality.


	Fig. \ref{PT_number} compares the total utility gain of different algorithms as the PT number increases. 
	The result shows the total utility gain decrease as the number of PT rises, with WDRO exhibiting a significantly slower rate of decline compared to the other algorithms.
	This is because, as the number of PTs increases while the overall network size and structure remain unchanged, storage limitations begin to become a major constraint. Once the PT count exceeds a certain threshold, the available storage capacity is no longer sufficient to deploy DT models at optimal locations. However, WDRO's inherent robustness enables it to better mitigate the adverse effects of these environmental changes, outperforming other algorithms.

	Fig. \ref{storage_capacity} shows the total utility gain as the maximum storage capacity of ES increases, with the components of utility gain from DT model update and service response latency. 
	The results indicate that WDRO outperforms in both utility gain of DT model update and service response latency. The utility gain rises and then plateaus as storage capacity increases, because further deployment of DT models no longer contributes significantly to the utility gain once the threshold is reached, which is sufficient to satisfy DT interaction requests. Additionally, WDRO achieves utility gain from service response similar to the Near-RQ, and its utility from DT model AoI close to the Near-PT, demonstrating its effectiveness in fulfilling requests and maintaining high-fidelity DT models.
	\vspace{-0.3em}

	\section{Conclusion}\label{Conclusion}
	In this paper, a joint optimization of DT model deployment and DT model selection for DTaaS provisioning over edge computing has been studied. To evaluate the quality of DTaaS, we introduce utility gain based on the AoI difference between edge and cloud service provisioning. With storage resource constraints, our goal is to maximize the total utility gain of all DT interaction requests. Focusing on providing robust solutions capable of adapting to unforeseen extreme request conditions, we propose a DRO-based approach, called WDRO, that reformulates the problem employing the DRO method. Then, by leveraging multi-level dual transformations based on Wasserstein distance, we derive a robust solution. Compared to counterparts, the proposed WDRO demonstrates the superiority in obtaining the total utility gain of all DT interaction requests under unforeseen extreme request conditions.
	\vspace{-0.2em}
	\section{Acknowledgments}
	This work was supported by State Key Laboratory of Massive Personalized Customization System and Technology No.H\&C-MPC-2023-04-01, and Postgraduate Research \& Practice Innovation Program of Jiangsu Province
	No.KYCX24\_0596.
	
	\bibliographystyle{IEEEtran}
	\bibliography{PaperRef.bib}

\end{document}